\documentclass{article}

\usepackage{PRIMEarxiv}

\usepackage[utf8]{inputenc} % allow utf-8 input
\usepackage[T1]{fontenc}    % use 8-bit T1 fonts
\usepackage{hyperref}       % hyperlinks
\usepackage{url}            % simple URL typesetting
\usepackage{booktabs}       % professional-quality tables
\usepackage{amsmath} 
\usepackage{amsfonts}       % blackboard math symbols
\usepackage{nicefrac}       % compact symbols for 1/2, etc.
\usepackage{microtype}      % microtypography
\usepackage{lipsum}
\usepackage{fancyhdr}       % header
\usepackage{graphicx}       % graphics
\graphicspath{{media/}}     % organize your images and other figures under media/ folder

\title{Radial Excitation Spectra of Light Pseudoscalar and Vector Mesons in Light-Front Holographic QCD
}

\author{Abhisth Srivastava$^{\dagger}$, \\
  Department of Physics,  \\
  Dr. B.R. Ambedkar National Institute of Technology, \\
  Jalandhar, Punjab-144011, India\\
  \\
  $^{\dagger}$\texttt{abhisth.srivastava16@gmail.com} \\
   \\
}

\begin{document}
\maketitle

\begin{abstract}
The non-perturbative regime of Quantum Chromodynamics (QCD) poses significant challenges for analytical treatment. Light-Front Holographic QCD (LFHQCD) offers an effective semiclassical framework for exploring hadronic structure through a holographic dual description of confinement. While the standard soft-wall model, along with the Brodsky–de Téramond (BdT) longitudinal prescription, provides satisfactory results for meson ground states, it shows deviations of around 17 to $18\%$ for radial excitations of the pion family. In this work, we extend the transverse confining potential by adding Coulombic and logarithmic contributions beyond the standard soft-wall framework. We solve the resulting light-front Schrödinger equation numerically, and we combine the transverse mass spectrum with the BdT longitudinal prescription to obtain the complete meson mass spectrum. We compute the radial excitation spectra of the $\pi$, $\rho$, and $K^{*}$ meson families and analyze their Regge trajectories. The modified framework reduces the overall average deviation from experimental data from about $7.6\%$ to $5.2\%$ compared to the standard soft-wall model, representing a $\sim 32\%$ improvement for the fitted nine-state dataset in describing radially excited states. The results show that short-range Coulombic and intermediate-range logarithmic contributions are important for refining the radial meson spectrum within the light-front holographic approach.
\end{abstract}

% keywords can be removed

\section{Introduction}\label{sec1}

Quantum Chromodynamics (QCD) is the SU(3) Yang-Mills gauge theory that describes strong interactions. It provides the basic understanding of quark and gluon dynamics. At high momentum transfer, asymptotic freedom makes the coupling constant small. This allows for perturbative calculations of high-energy processes, such as deep inelastic scattering and jet production. However, in the low-energy region of QCD, we encounter color confinement, chiral symmetry breaking, and hadron spectroscopy. These aspects fall into the nonperturbative domain. The strong coupling in this area makes it very difficult to treat analytically. One major unresolved challenge in theoretical physics is finding a first-principles description of hadronic bound states.

Light-Front Quantization gives a straightforward Hamiltonian framework for describing the internal structure of hadrons using constituent parton degrees of freedom \cite{ MatthiasBurkardt1995, Brodsky1998}. In this approach, hadronic observables are expressed directly through light-front wave functions. This offers a frame-independent and boost-invariant description of bound states. Building on this foundation, Light-Front Holographic QCD (LFHQCD), developed by Brodsky and de Téramond \cite{Brodsky2006, deTeramond2005, Brodsky2014}, reveals a clear link between the light-front Schrödinger equation that governs hadronic dynamics and the equations of motion in Anti-de Sitter (AdS) space. This connection comes from the AdS/CFT duality proposed by Maldacena \cite{Maldacena1999}. It provides an effective semiclassical description of confinement and hadronic structure without needing explicit perturbative input.

The soft-wall model within the LFHQCD framework introduces a quadratic dilaton profile. This creates a harmonic confining potential in the transverse direction \cite{deTeramond2009, Brodsky2015, Brodsky2011}. This setup matches observations, reproducing linear Regge trajectories and providing analytical mass formulas for light mesons and baryons. Researchers have explored extensions of the soft-wall framework, which include changes to the dilaton profile and the addition of quark mass effects, to better match experimental data \cite{Guo2016, Travinski2014, Swarnkar2015}. The longitudinal dynamics of hadronic bound states, including the impact of finite quark masses, have been incorporated through the Brodsky–de Téramond (BdT) prescription \cite{Ahmady2021, AhmadyHarleen2021}. This approach yields satisfactory results for the ground state masses of light mesons. Additionally, meson transition form factors and diffractive production processes have been examined within this holographic framework \cite{Brodsky2011, Ahmady2016}, showing its wide applicability to hadronic observables.

Despite these achievements, the standard soft-wall model has notable limitations in capturing radially excited states. In this study, we show that the soft-wall framework, when combined with the BdT longitudinal prescription, results in discrepancies of about 17–$18\%$ for the radial excitations of the pion family. This is noticeably higher than the deviations for ground states. This finding calls for a systematic improvement of the transverse confining potential beyond the simple harmonic soft-wall form. Phenomenological studies indicate that contributions from one-gluon exchange and logarithmic terms that reflect intermediate-range nonperturbative dynamics can significantly enhance hadronic mass predictions \cite{Pandya2024, Travinski2014}. The radial structure of Regge trajectories and the specific dependence of meson masses on the radial quantum number impose strict constraints on such enhancements \cite{Masjuan2012}.

Multiple recent works have extended light-front holographic QCD with complementary longitudinal dynamics or additional short-range structure for specific meson sectors, including the pion \cite{Ahmady2023}, the $\rho$ meson \cite{Gurjar2024}, the $\phi$ meson \cite{Gurjar2025}, and radiative transition form factors for light mesons \cite{Ahmady2020, Ahmady2018}. The current study follows this direction by incorporating Coulombic and logarithmic contributions into the transverse dynamics of the $\pi$, $\rho$, and $K^{*}$ radial trajectories, by adapting these terms, originally introduced in a heavy-meson light-front quark model \cite{Pandya2024} as part of a linear mass eigenvalue equation, to the present mass-squared framework while preserving dimensional consistency (Sec. 3). We solve the resulting light-front Schrödinger equation numerically and combine the transverse mass spectrum with the Brodsky–de Téramond (BdT) longitudinal prescription to obtain the complete meson mass spectrum. We first establish a baseline by computing the radial excitation spectra of the $\pi$, $\rho$, and $K^{*}$ meson families within the conventional soft-wall framework, and then show that the modified potential consistently reduces the average deviation from experimental data \cite{Workman2022}, from about $7.6\%$ to $5.2\%$, a $\sim 32\%$ relative improvement for the fitted nine-state dataset in describing radially excited states.

This paper is structured as follows. Section~\ref{sec2} recaps the LFHQCD framework. Section~\ref{sec3} explains the extended transverse potential and its physical justification, including the dimensional-consistency argument underlying the Coulombic and logarithmic terms. Section~\ref{sec4} presents the numerical procedure and the resulting mass spectra, together with the respective Regge trajectories. Section~\ref{sec5} compares the modified framework against the baseline soft-wall model. Section~\ref{sec6} discusses the physical significance of the results. Section~\ref{sec7} concludes with suggestions for future research. The appendix~\ref {secA1} details the numerical procedure.

\section{Light-Front Holographic QCD}\label{sec2}
The invariant mass of a relativistic $q\bar{q}$ bound state in (3+1)-dimensional Light-Front QCD is obtained from the eigenvalue equation of the light-front Hamiltonian. For a two-body mesonic system, the squared mass operator can be written as (\cite{Brodsky2015}) (\cite{Ahmady2021}),  

\begin{equation} \label{Eq:1}
M^2 = \int \mathrm{d}x \, \mathrm{d}^2 \mathbf{b}_\perp \, \Psi^*(x, \mathbf{b}_\perp) 
\left[ -\frac{\nabla_{b_\perp}^2}{x(1-x)} + \frac{m_q^2}{x} + \frac{m_q^2}{1-x} \right]
\Psi(x, \mathbf{b}_\perp) + \text{interactions}
\end{equation}

where $x$ is the quark light-front momentum fraction, $\mathbf{b}_\perp$ ($\mathbf{b}_\perp = b_\perp e^{i\varphi}$) is the transverse distance between quark and antiquark, $\Psi(x, b_\perp)$ is the light-front wavefunction, and $m_q$ and $m_{\bar{q}}$ are the quark and antiquark masses.

Using light-front variable, 
\begin{equation} \label{Eq:2}
\boldsymbol{\zeta} = \sqrt{x(1-x)} \mathbf{b}_\perp
\end{equation}
the wave function is resolved into transversed and longitudinal form given by, 

\begin{equation} \label{Eq:3}
    \Psi (x,\zeta, \varphi)= \frac{\phi(\zeta)}{\sqrt{2\pi\zeta}} e^{iL\varphi} X(x)
\end{equation}
\\where $\phi(\zeta)$ and $X(x)=\sqrt{x(1-x)} \chi(x)$ are the transverse and longitudinal modes respectively. L is light-front orbital angular momentum (\cite{Brodsky2010}).

Substituting this factorized form into the Hamiltonian expression yields a separation of the mass operator into transverse and longitudinal contributions,

\begin{equation} \label{Eq:4}
    M^2= M^2_\perp + M^2_\parallel
\end{equation}
\\where, 
\begin{equation} \label{Eq:5}
    M^2_\perp= \int d^2\boldsymbol{\zeta} \phi^* (\zeta) \left  [-\frac{d^2}{d\zeta^2} + \frac{4L^2-1}{4\zeta^2}+ U_\perp(\zeta)\right]
\end{equation}

Subject to the normalization condition,
\begin{equation} \label{Eq:6}
    \int dx |\chi(x)|^2 = 1
\end{equation}

In Anti de-Sitter ($AdS_5$) space, due to holographic mapping in the same, the Transverse Confinement Potential is (\cite{Travinski2014}) (\cite{Ahmady2016}) (\cite{deTeramond2010}), 
\begin{equation} \label{Eq:7}
    U_\perp(\zeta)= \kappa^4\zeta^2+2\kappa^2(J-1)
\end{equation}
where J = \textit{L} + \textit{S}. Eq. (\ref{Eq:5}) can be rewritten as (\cite{Brodsky2015}) (\cite{Ahmady2021}) (\cite{deTeramond2009}),
\begin{equation} \label{Eq:8}
    \left(-\frac{d^2}{d\zeta^2} + \frac{4L^2-1}{4\zeta^2}+ U_\perp(\zeta)\right)\phi(\zeta)=M^2_\perp \phi(\zeta).
\end{equation}

\section{Transverse Potential Parameterization}\label{sec3}

The light-front holographic approach provides an effective framework for probing hadronic 
structure by mapping the constraints of quantum chromodynamics (QCD) onto a semiclassical 
light-front Schr\"odinger equation. In this formulation, confinement is applied through an 
effective transverse potential that controls the dynamics of quark--antiquark bound states. 
The standard soft-wall model systematically reproduces the linear Regge behavior observed in 
light meson spectra through a harmonic confining interaction, while maintaining analytic 
agreement.

In its standard form, the transverse light-front potential consists of a quadratic confinement 
term supported by a spin-dependent contribution. This structure captures the crucial 
long-distance dynamics responsible for confinement. However, it does not fully account for 
short- and intermediate-range effects arising from gluon exchange and other nonperturbative 
mechanisms. To account for such effects, we consider an extended transverse potential of the 
form \cite{Pandya2024},
\begin{equation}
U_{\perp}(\zeta) = \kappa^4 \zeta^2 + 2\kappa^2(J - 1) - \frac{4\alpha_C \kappa}{3\zeta} 
+ c_{\text{eff}} \ln\left(\frac{\zeta}{\zeta_0}\right)
\label{Eq:9}
\end{equation}
where $\zeta$ denotes the invariant transverse separation between the quark and antiquark, 
$\kappa$ is the confinement scale, and $J=L+S$ represents the total light-front angular 
momentum. The Coulombic term proportional to $\alpha_C$ accounts for short-range interactions 
associated with one-gluon exchange, while the logarithmic contribution introduces an 
intermediate-range modification motivated by phenomenological considerations. We denote this parameter $\alpha_C$ rather than the conventional $\alpha_s$ to emphasize that 
it is a phenomenological fit parameter to the light-meson spectrum, not the QCD running 
coupling evaluated at a specified renormalization scale. Together, these 
terms extend the standard soft-wall potential while preserving its confining nature.

The Coulomb and logarithmic terms are adapted from the instant-form potential of 
Pandya et al.~\cite{Pandya2024}, where they enter a mass-linear eigenvalue equation, 
$H|\Psi\rangle = M|\Psi\rangle$. Since the present transverse equation is instead an eigenvalue 
equation for $M^2$ (Eq.~\ref{Eq:8}), a direct substitution is not dimensionally consistent. 
Following the general instant-form/front-form correspondence of 
Trawi\'nski et al.~\cite{Travinski2014}, an instant-form potential $V_{\rm eff}$ maps onto an 
effective front-form potential $U_{\rm eff}\approx 2M\,V_{\rm eff}$ at leading order in 
$V_{\rm eff}$.

For the Coulomb contribution, the precise normalization of the promotion scale is not 
independently identifiable because $\alpha_C$ is fitted to the meson spectrum. Replacing 
$\kappa$ by any fixed multiple $a\kappa$ is exactly equivalent, after refitting $\alpha_C$, to 
redefining $\alpha_C\rightarrow a\,\alpha_C$, and therefore leaves the fitted spectrum 
unchanged. This ambiguity is thus a parametrization convention rather than an independent 
physical assumption.

The logarithmic term differs because its coefficient $c_{\rm eff}=c_{QR}\kappa$ is fixed 
rather than independently fitted. We therefore adopt $\kappa$, the intrinsic confinement scale 
of the soft-wall framework, as the representative fixed scale. This choice is motivated by two 
considerations. First, $c_{QR}$ originates from the phenomenological Quigg--Rosner logarithmic 
potential~\cite{Quigg1977}, rather than being a first-principles QCD coefficient. Second, 
retaining the literal state-dependent mass $M$ in $U_{\rm eff}\approx 2M V_{\rm eff}$ would 
place $M$ simultaneously in the potential and in the eigenvalue, resulting in a self-consistent 
nonlinear eigenvalue problem rather than the linear one considered here. The resulting 
prescription is therefore a phenomenological implementation of the instant-form/front-form 
correspondence, rather than a unique first-principles derivation.

The logarithmic contribution is quantitatively significant: its expectation value exceeds that 
of the Coulomb contribution for all states considered, with 
$\langle V_{\log}\rangle/\langle V_C\rangle$ increasing from approximately $1.6$ at $n=0$ to 
$6.6$ at $n=2$. We therefore explicitly assess the model dependence associated with its 
promotion scale by taking $c_{\rm eff}=\lambda c_{QR}\kappa$ and independently refitting 
$\alpha_C$ at each value of $\lambda$, thereby isolating the sensitivity to the logarithmic 
normalization rather than conflating it with the freedom already present in $\alpha_C$. Over 
$0.5\leq\lambda\leq2$, the mean deviation remains between $4.21\%$ and $6.09\%$ 
(Table~\ref{tab:sensitivity}), indicating moderate model dependence. Although the mean 
deviation is lower at $\lambda=1.5$ than at the adopted $\lambda=1$, this improvement is not 
uniform: it comes at the cost of substantially worse agreement in the $K^*$ sector 
($K^*(1410)$: $4.3\%\to8.1\%$; $K^*(1680)$: $9.3\%\to12.7\%$), while improving the $\pi$ and 
$\rho$ excited states. We therefore retain $\lambda=1$ as the physically motivated reference 
choice associated with the intrinsic confinement scale, rather than selecting $\lambda$ solely 
by minimizing the global mean deviation.

\begin{table}[h]
\centering
\caption{Sensitivity of the fit to the logarithmic promotion scale, 
$c_{\rm eff}=\lambda\,c_{QR}\kappa$. $\alpha_C$ is independently refit at each $\lambda$; 
mean deviation is averaged over the same nine PDG states as Table~\ref{tab:baseline}.}
\label{tab:sensitivity}
\begin{tabular}{ccc}
\hline\hline
$\lambda$ & Refit $\alpha_C$ & Mean deviation \\
\hline
0.5 & 0.109 & 6.09\% \\
1.0 (adopted) & 0.172 & 5.16\% \\
1.5 & 0.206 & 4.21\% \\
2.0 & 0.220 & 4.31\% \\
\hline\hline
\end{tabular}
\end{table}

We take $\kappa=0.523$ GeV \cite{Ahmady2021,Brodsky2015}, $c_{QR}=0.733$ GeV and 
$\zeta_0=0.89$ GeV$^{-1}$, the values adopted from Quigg and Rosner~\cite{Quigg1977} as used by 
Pandya et al.~\cite{Pandya2024}, giving $c_{\text{eff}}=c_{QR}\,\kappa=0.383$ GeV$^2$. 
$\alpha_C$ is fit to the light-meson spectrum in Sec.~4, giving $\alpha_C=0.172$.

We note that the pion's status as a pseudo-Goldstone boson of spontaneously broken chiral 
symmetry introduces dynamics not explicitly built into this extension 
\cite{Choi2015,Li2022,Roberts2021,Ballon-Bayona2015,Brodsky2008}; this may contribute to the 
larger residual deviation seen in the pion's excited states Sec.\ref{sec5}.

In the original soft-wall formulation, the transverse light-front Schr\"odinger equation admits 
an analytic solution, yielding the well-known mass spectrum \cite{Ahmady2021,Brodsky2015},
\begin{equation}
M^2 = 4\kappa^2\left(n + L + \frac{S}{2}\right),
\label{Eq:10}
\end{equation}
where $n$ denotes the radial quantum number. This expression leads to linear Regge trajectories 
with a universal slope determined by the confinement scale $\kappa$. The soft-wall model is 
conformal at the classical level. The addition of Coulombic and logarithmic contributions 
introduces explicit breaking of conformal symmetry. The current analysis does not aim to keep 
superconformal symmetry \cite{Guo2016,deTeramond2005}. Instead, it focuses on improving the 
radial spectrum.

Substituting the extended potential of Eq.~\ref{Eq:9} into Eq.~\ref{Eq:8}, the transverse 
eigenvalue equation becomes
\begin{equation}
\left[-\frac{d^2}{d\zeta^2} + \frac{4L^2-1}{4\zeta^2} + \kappa^4\zeta^2 + 2\kappa^2(J-1) 
- \frac{4\alpha_C\kappa}{3\zeta} + c_{\text{eff}}\ln\!\left(\frac{\zeta}{\zeta_0}\right)\right]
\phi(\zeta) = M_\perp^2\,\phi(\zeta),
\label{Eq:9b}
\end{equation}
subject to $\phi(0)=\phi(\infty)=0$ and $\int_0^\infty|\phi(\zeta)|^2\,d\zeta=1$. Unlike the 
pure soft-wall case (Eq.~\ref{Eq:10}), the Coulomb and logarithmic terms preclude an analytic 
solution, and Eq.~\ref{Eq:9b} is solved numerically as described in Sec. \ref{sec4}.

The inclusion of Coulombic and logarithmic contributions, however, renders the eigenvalue 
problem analytically intractable. In this case, the transverse mass spectrum must be obtained 
numerically. Despite this modification, the resulting eigenvalues continue to show an 
approximately linear dependence on the radial quantum number, allowing the spectrum to be 
configured in the form,
\begin{equation}
M_\perp^2 = a \left(n + L + \frac{S}{2}\right),
\label{Eq:11}
\end{equation}
where the coefficient $a$ represents an effective Regge slope incorporating the effects of the 
modified potential. We note that, unlike the exact soft-wall spectrum, the extended potential 
does not yield perfectly linear $M_\perp^2$ versus $(n+L+S/2)$; we report the proportional fit 
$a=1.2519$ GeV$^2$ (zero intercept, for direct comparison with Eq.~\ref{Eq:10}) as an effective 
slope, while the mass predictions used throughout this work (Tables~1--3) are computed from the 
full numerical eigenvalues rather than reconstructed from this fit.

This methodology provides a consistent framework for examining the transverse structure of 
light mesons while allowing controlled deviations from the pure soft-wall limit. In the 
subsequent sections, we apply this framework to compute the mass spectra of light pseudoscalar 
and vector mesons and evaluate the impact of the modified potential on their Regge trajectories.

\section{Numerical Result and Spectroscopy}\label{sec4}

\subsection{Numerical Method}

We solve the transverse light-front Schr\"odinger equation, Eq.~(\ref{Eq:9b}), by 
diagonalizing the Hamiltonian in the basis of exact soft-wall eigenfunctions, 
$\phi_n(\zeta)=\kappa\sqrt{2}\,\zeta^{1/2}e^{-\kappa^2\zeta^2/2}L_n(\kappa^2\zeta^2)$, which 
satisfy the unperturbed ($\alpha_C=c_{\rm eff}=0$) equation exactly, with eigenvalues 
$2\kappa^2(2n+1)$ for $L=0$. Matrix elements of the Coulomb and logarithmic perturbation are 
computed by numerical quadrature and the resulting matrix is diagonalized to obtain 
$M_\perp^2$. The method was validated against the exact analytic soft-wall spectrum, 
Eq.~(\ref{Eq:10}), to machine precision before adding the extension. Convergence with respect 
to basis size was checked explicitly; Table~\ref{tab:convergence} shows the lowest three 
eigenvalues of the extended potential for the vector ($J=1$) sector, the corresponding masses change by less than 0.4 MeV between 25 and 45 basis states. All results reported below use 25 basis states. Details 
are provided in Appendix~\ref{secA1}.

\begin{table}[h]
\centering
\caption{Convergence of the lowest three $M_\perp^2$ eigenvalues (GeV$^2$) of the extended 
potential, vector ($J=1$) sector, with basis size.}
\label{tab:convergence}
\begin{tabular}{cccc}
\hline\hline
$N_{\rm basis}$ & $n=0$ & $n=1$ & $n=2$ \\
\hline
10 & 0.548958 & 1.922249 & 3.131703 \\
15 & 0.547984 & 1.921431 & 3.130745 \\
20 & 0.547579 & 1.921119 & 3.130412 \\
25 & 0.547363 & 1.920959 & 3.130250 \\
30 & 0.547231 & 1.920865 & 3.130155 \\
35 & 0.547142 & 1.920802 & 3.130094 \\
45 & 0.547033 & 1.920727 & 3.130021 \\
\hline\hline
\end{tabular}
\end{table}

\subsection{Effective Regge Slope}

Despite the modification, the resulting eigenvalues continue to show an approximately linear 
dependence on the radial quantum number, allowing the spectrum to be summarized, for comparison 
with Eq.~(\ref{Eq:10}), in the form of Eq.~(\ref{Eq:11}), with a fitted proportional slope
\begin{equation} \label{Eq:12}
    a = 1.2519 \ \text{GeV}^2 .
\end{equation}
We note that this is an approximate summary slope, obtained by a zero-intercept fit to the 
computed eigenvalues, and not the calculation itself; the extended potential does not yield 
perfectly linear $M_\perp^2$ versus $(n+L+S/2)$ (see Sec.~\ref{sec5}). The mass predictions used 
throughout this work (Tables~3--5) are computed from the full numerical eigenvalues of 
Eq.~(\ref{Eq:9b}), rather than reconstructed from Eq.~(\ref{Eq:12}).

\subsection{Longitudinal Correction and Fitting Procedure}

Taking into account the effect of non-zero quark masses, we used the prescription developed by 
Brodsky and de T{\'e}ramond (BdT) \cite{Ahmady2021,Brodsky2009,AhmadyHarleen2021} for pion mass, 
given by 
\begin{equation} \label{Eq:13}
    \Delta M_{BdT}^{2} = \frac{\int_{0}^{1} dx\, \chi_{BdT}^{2}(x) \left( \frac{m_{q}^{2}}{x} + \frac{m_{\bar{q}}^{2}}{1-x} \right)}{\int_{0}^{1} dx\, \chi_{BdT}^{2}(x)}
\end{equation}
where, 
\begin{equation} \label{Eq:14}
    \chi_{BdT} = \exp\left(-\frac{(1-x)m_q^2 + xm_{\bar{q}}^2}{2\kappa^2x(1-x)}\right)
\end{equation}
The given equation is the normalised expression for BdT.
It is to be noted that Eq.~(\ref{Eq:13}) is accurate only for light hadrons in their ground 
states, with satisfactory results shown for excited states too. This also accounts for linear 
Regge behaviour. The transverse dynamics are taken to be flavor-independent: the same 
$\alpha_C$, $\kappa$, $c_{\rm eff}$, and $\zeta_0$ are used for the $\pi$, $\rho$, and $K^{*}$ 
sectors, with the strange-sector mass splitting generated entirely by the quark-mass dependence 
of $\Delta M_{\rm BdT}^2$ in Eq.~(\ref{Eq:13}). With $\kappa$, $c_{\rm eff}$, and $\zeta_0$ fixed 
as in Table~\ref{tab:parameters}, $\alpha_C$ is the only free parameter. It is determined by a 
bounded scalar minimization of the mean absolute relative deviation, 
$\frac{1}{9}\sum_i|M_i^{\rm calc}-M_i^{\rm PDG}|/M_i^{\rm PDG}$, over the nine states listed in 
Tables 3--5, with $\alpha_C\in[0,2]$. We solved Eq.~(\ref{Eq:13}), together with 
Eq.~(\ref{Eq:9b}), to fit the light-hadron spectrum, using the universal $\kappa=0.523$ GeV together with the mass of up and down quark as $m_{u/d}$ = 0.046 GeV and mass of strange quark $m_{s}$ = 0.357 GeV, giving $\alpha_C = 0.172$. We do not propagate the quoted uncertainty on 
$\kappa$ ($\pm0.024$ GeV \cite{Brodsky2015}) into the mass predictions below; doing so is left 
for future work.

\begin{table}[h]
\centering
\caption{Input parameters used in the numerical calculation. 
The logarithmic coefficient $c$ is adapted from the $r$-space 
value of Quigg and Rosner (\cite{Quigg1977}) as 
$c_{\rm eff} = c_{\rm QR} \times \kappa$, and the Coulomb term 
is similarly rescaled by $\kappa$, to ensure dimensional 
consistency in the $\zeta$-space mass-squared equation 
(see Sec.~3). $\alpha_C$ is fit to the light-meson spectrum 
(Sec.~4.3); the BdT corrections are evaluated with the explicit 
normalization of Eq.~(\ref{Eq:13}).}
\label{tab:parameters}
\begin{tabular}{lllll}
\hline\hline
Parameter & Symbol & Value & Unit & Source\\
\hline
Confinement scale         
    & $\kappa$                      
    & $0.523$     
    & GeV           
    & \cite{Brodsky2015} \\
Up/down quark mass        
    & $m_{u/d}$                     
    & $0.046$               
    & GeV           
    & \cite{Ahmady2021} \\
Strange quark mass        
    & $m_s$                         
    & $0.357$               
    & GeV           
    & \cite{Ahmady2021} \\
Coulomb strength          
    & $\alpha_C$                    
    & $0.172$               
    & --            
    & This work (fit) \\
Log.\ coefficient ($r$-space)    
    & $c_{\rm QR}$                  
    & $0.733$               
    & GeV           
    & \cite{Quigg1977,Pandya2024} \\
Log.\ coefficient ($\zeta$-space) 
    & $c_{\rm eff} = c_{\rm QR}\,\kappa$ 
    & $0.383$               
    & GeV$^2$       
    & This work \\
Logarithmic scale         
    & $\zeta_0$                     
    & $0.890$               
    & GeV$^{-1}$    
    & \cite{Quigg1977,Pandya2024} \\
BdT mass correction ($u/d$)      
    & $\Delta M^2_{\rm BdT}(\pi,\rho)$        
    & $0.0194$               
    & GeV$^2$       
    & This work \\
BdT mass correction ($u/d$-$s$)  
    & $\Delta M^2_{\rm BdT}(K^{*})$  
    & $0.241$               
    & GeV$^2$       
    & This work \\
\hline\hline
\end{tabular}
\end{table}

\subsection{Mass Spectra}

The full meson mass is then given by (\cite{Ahmady2021}),
\begin{equation} \label{Eq:15}
M^{2} = M_{\perp}^{2} (n, L, S) + \Delta M_{\mathrm{BdT}}^{2} 
\end{equation}
where \(M_{\perp}^{2}\) represents the transverse contribution obtained from the numerical 
solution of the light-front Schr\"odinger equation, and \(\Delta M_{\mathrm{BdT}}^{2}\) denotes 
the longitudinal mass correction following the Brodsky--de T\'eramond prescription. 

The BdT mass evaluated for the up-down quark system is 0.0194 GeV$^2$ and for the 
up/down-strange quark system is 0.241 GeV$^2$.
    
 The Parity is given by, 
 \begin{equation} \label{Eq:16}
     P=(-1)^{L+1}
 \end{equation}
And the charge-conjugation quantum number is given by, 
\begin{equation} \label{Eq:17}
    C=(-1)^{L+S}
\end{equation}

We note that $C$ is a well-defined quantum number only for flavor-neutral, self-conjugate 
states, and is therefore reported only for the $\pi$ and $\rho$ families; the $K^{*}$ mesons 
carry net strangeness and are consequently classified by $J^P$ alone. Using Eqs.~(\ref{Eq:9b}), (\ref{Eq:13}), (\ref{Eq:15}), (\ref{Eq:16}) and (\ref{Eq:17}), we 
computed spectra for pseudoscalar mesons (L=0, S=0) and vector mesons (L=0, S=1), compared 
directly with PDG data \cite{Workman2022}.

\begin{table}[h]
\centering
\large
\begin{tabular}{|c|c|c|c|c|c|c|}
\hline
$J^{PC}$ & Name & $M_\parallel$ & $M_\perp$ & $M$ & PDG & Dev. \\
\hline
$0^{-+}$ & $\pi(140)$ & 139 & 17 & 140 & 140 & 0.00\% \\
\hline
$0^{-+}$ & $\pi(1300)$ & 139 & 1172 & 1180 & 1300 & 9.23\% \\
\hline
$0^{-+}$ & $\pi(1800)$ & 139 & 1607 & 1613 & 1800 & 10.39\% \\
\hline
\end{tabular}
\caption{Computed masses of pion and their radial excitations (in MeV) using mixed potential, 
compared with PDG data \cite{Workman2022}.}
\label{Tab3}
\end{table}

\begin{table}[h] 
\centering
\large
\begin{tabular}{|c|c|c|c|c|c|c|}
\hline
$J^{PC}$ & Name & $M_\parallel$ & $M_\perp$ & $M$ & PDG & Dev. \\
\hline
$1^{--}$ & $\rho(770)$ & 139 & 740 & 753 & 770 & 2.21\% \\
\hline
$1^{--}$ & $\rho(1450)$ & 139 & 1386 & 1393 & 1450 & 3.93\% \\
\hline
$1^{--}$ & $\rho(1900)$ & 139 & 1769 & 1775 & 1900 & 6.58\% \\
\hline
\end{tabular}
\caption{Computed masses of $\rho$ meson and their radial excitations (in MeV) using mixed 
potential, compared with PDG data \cite{Workman2022}.}
\label{Tab4}
\end{table}

\begin{table}[h]
\centering
\large
\begin{tabular}{|c|c|c|c|c|c|c|}
\hline
$J^{PC}$ & Name & $M_\parallel$ & $M_\perp$ & $M$ & PDG & Dev. \\
\hline
$1^{-}$ & $K^{*}(892)$ & 491 & 740 & 888 & 892 & 0.45\% \\
\hline
$1^{-}$ & $K^{*}(1410)$ & 491 & 1386 & 1470 & 1410 & 4.26\% \\
\hline
$1^{-}$ & $K^{*}(1680)$ & 491 & 1769 & 1836 & 1680 & 9.29\% \\
\hline
\end{tabular}
\caption{Computed masses of K* meson and their radial excitations (in MeV) using mixed 
potential, compared with PDG data \cite{Workman2022}.}
\label{Tab5}
\end{table}

In this work we focused on the radial excitations of light mesons; therefore, we analyzed the 
dependence of the squared mass on the radial quantum number n while keeping the orbital angular 
momentum L fixed.

\begin{figure}
    \centering
    \includegraphics[width=0.75\linewidth]{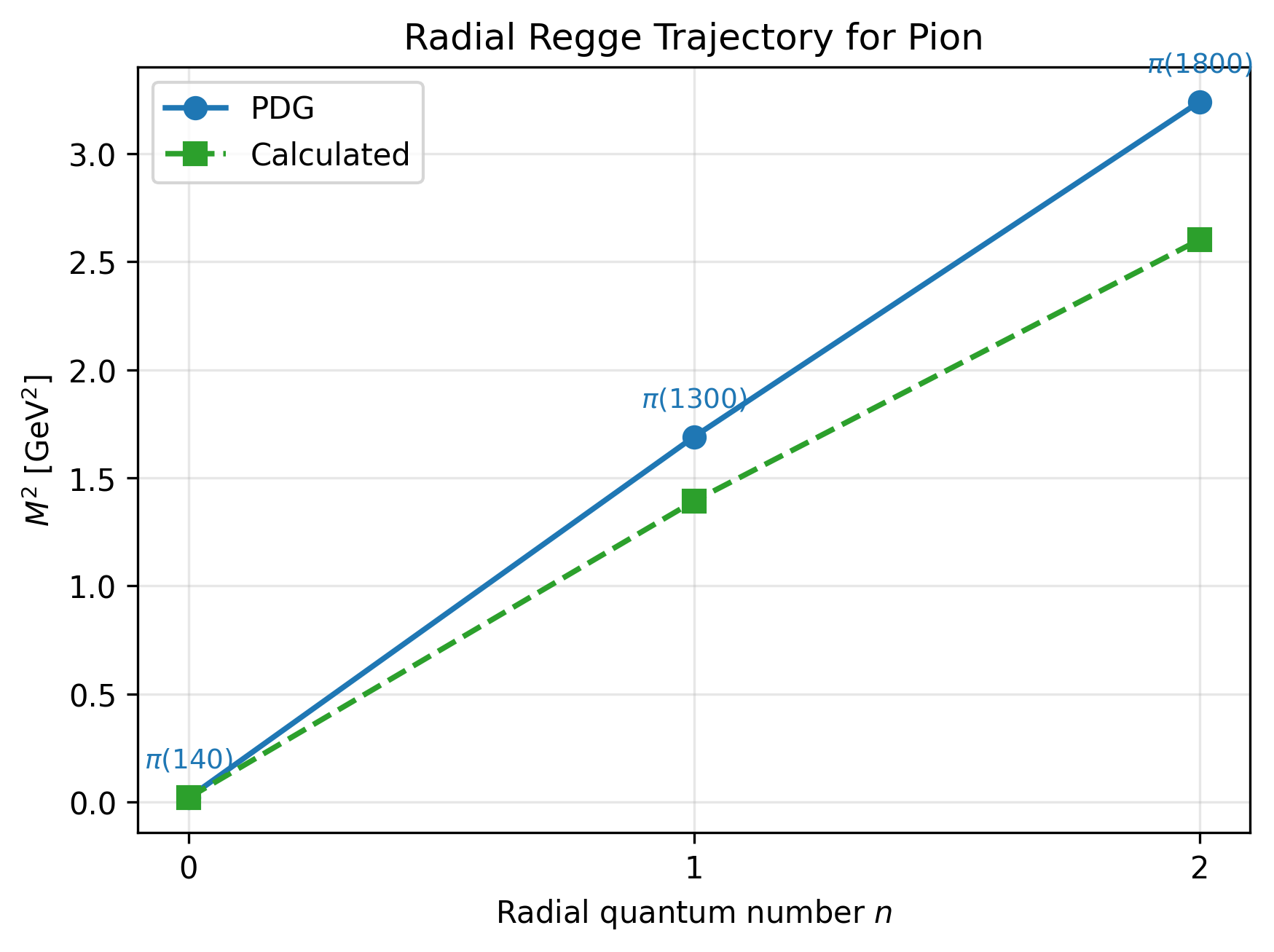}
    \caption{Regge Trajectory of Pion family. The Blue curve is defined for the PDG value of Pions, whereas Green curve is defined for calculated values of Pions.\cite{Workman2022}}
\end{figure}

\begin{figure}
    \centering
    \includegraphics[width=0.75\linewidth]{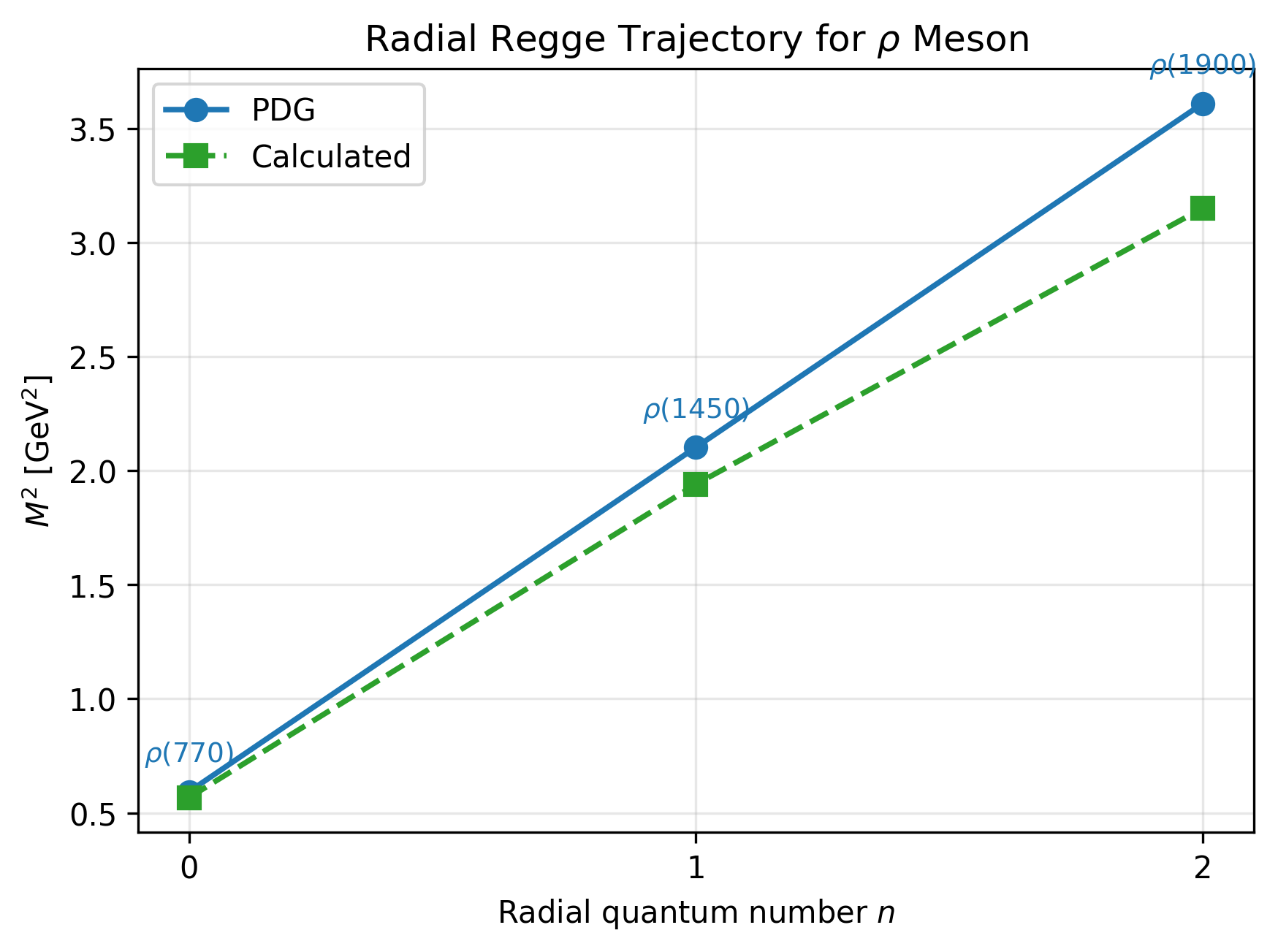}
    \caption{Regge Trajectory of Rho Meson family. The Blue curve is defined for the PDG value of Rho Mesons, whereas Green curve is defined for calculated values of Rho Mesons.\cite{Workman2022}}
\end{figure}

\begin{figure}
    \centering
    \includegraphics[width=0.75\linewidth]{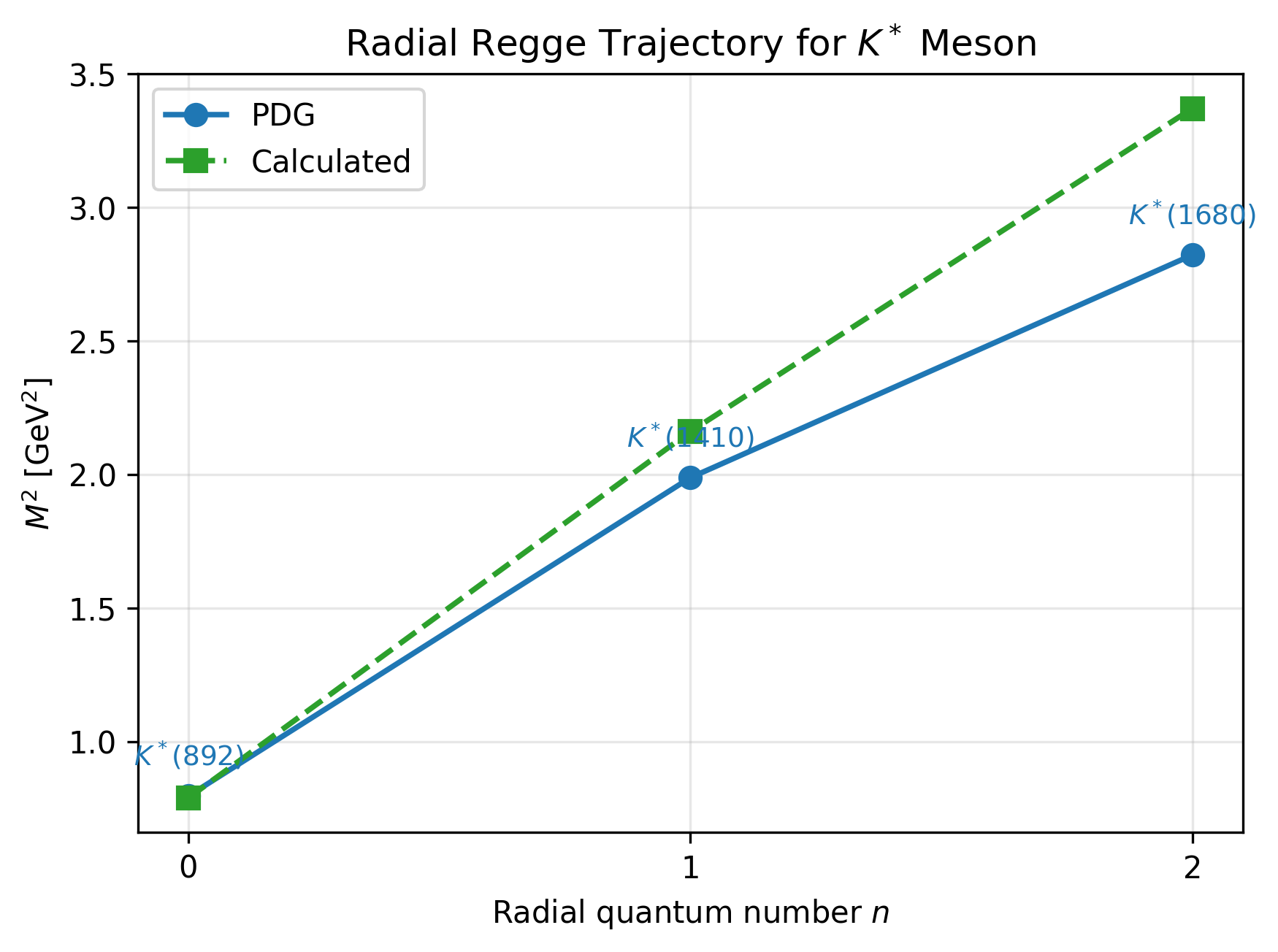}
    \caption{Regge Trajectory of $K^{*}$ Meson family. The Blue curve is defined for the PDG value of $K^{*}$ Mesons, whereas Green curve is defined for calculated values of $K^{*}$ Mesons.\cite{Workman2022}}
\end{figure}

\section{Comparison with the Baseline Soft-Wall Model}\label{sec5}

To assess whether the Coulombic and logarithmic contributions genuinely improve the 
description of the light-meson spectrum, rather than merely introducing additional free 
parameters, we compare the extended potential of Eq.~(\ref{Eq:9}) against the unmodified 
soft-wall model (Eq.~\ref{Eq:10}), using the same $\kappa$, quark masses, and normalized BdT 
longitudinal correction (Eq.~\ref{Eq:13}) in both cases. Table~\ref{tab:baseline} summarizes 
the resulting masses and their deviations from the PDG values \cite{Workman2022}.

\begin{table}[h]
\centering
\caption{Comparison of the unmodified soft-wall model (baseline) and the extended potential 
(this work) against PDG data \cite{Workman2022}. All masses in MeV.}
\label{tab:baseline}
\begin{tabular}{|c|c|c|c|c|c|}
\hline
State & PDG & Baseline & Dev. & Extended & Dev. \\
\hline
$\pi(140)$  & 140  & 139.4  & 0.43\%  & 140  & 0.00\% \\
$\pi(1300)$ & 1300 & 1055.2 & 18.83\% & 1180 & 9.23\% \\
$\pi(1800)$ & 1800 & 1485.8 & 17.46\% & 1613 & 10.39\% \\
\hline
$\rho(770)$  & 770  & 752.7  & 2.25\%  & 753  & 2.21\% \\
$\rho(1450)$ & 1450 & 1288.6 & 11.13\% & 1393 & 3.93\% \\
$\rho(1900)$ & 1900 & 1659.7 & 12.65\% & 1775 & 6.58\% \\
\hline
$K^{*}(892)$  & 892  & 887.8  & 0.47\% & 888  & 0.45\% \\
$K^{*}(1410)$ & 1410 & 1371.9 & 2.70\% & 1470 & 4.26\% \\
$K^{*}(1680)$ & 1680 & 1725.2 & 2.69\% & 1836 & 9.29\% \\
\hline
\end{tabular}
\end{table}

Averaged over all nine states, the mean deviation from PDG data decreases from $7.62\%$ for the 
unmodified soft-wall model to $5.15\%$ with the extended potential, a relative improvement of 
approximately $32\%$ evaluated on the same nine states used to fit $\alpha_C$. The improvement is concentrated almost entirely in the pion and $\rho$ 
sectors, where the baseline model underestimates the excited-state masses substantially 
(e.g., $\rho(1450)$ improves from $11.13\%$ to $3.93\%$ deviation). In the $K^{*}$ sector, by 
contrast, the extension slightly worsens the agreement for the higher radial excitations 
($K^{*}(1410)$: $2.70\%\to4.26\%$; $K^{*}(1680)$: $2.69\%\to9.29\%$), even though the ground state 
is essentially unaffected. This is a direct consequence of the flavor-independence assumption 
stated in Sec.~4.3: a single $\alpha_C$, fit globally across all nine states, cannot 
simultaneously optimize the light ($u/d$) and strange sectors, and the fit is dominated by the 
larger absolute improvements available in the pion and $\rho$ trajectories. A 
flavor-dependent refit of $\alpha_C$ -- or a $K^{*}$-specific extension of the potential is a 
natural direction for future work, and is left unexplored here in order to keep the number of 
free parameters to a minimum (Sec.~4.3).

Overall, the comparison indicates that the Coulombic and logarithmic contributions capture 
genuine short- and intermediate-range physics relevant to the radial excitation spectrum of 
light unflavored mesons, at the cost of a modest trade-off in the strange sector under the 
present flavor-independent treatment.

\section{Discussions}\label{sec6}

We computed the meson spectra in Sec.~\ref{sec4}, where the Regge trajectories for the pion, $\rho$, and $K^{*}$ families were presented, and compared the extended potential with the original soft-wall model in Sec.~\ref{sec5}. Averaged over the nine states considered, the mean deviation from the PDG masses \cite{Workman2022} decreases from $7.62\%$ for the baseline soft-wall model to $5.15\%$ with the extended potential, corresponding to an overall improvement of approximately $32\%$ representing an in-sample reduction in mean deviation rather than an independent out-of-sample test. The average deviations within the individual meson families are $6.54\%$ for the pion, $4.24\%$ for the $\rho$, and $4.67\%$ for the $K^{*}$ states. This improvement, however, is not uniform. The largest gains occur in the pion and $\rho$ sectors, whereas the higher excited $K^{*}$ states are described slightly less accurately than in the baseline model. As discussed in Sec.~\ref{sec5}, this reflects the use of a single flavor-independent value of $\alpha_C$ for all nine states. The sensitivity of the extended potential to the logarithmic promotion scale is quantified in 
Sec.~\ref{sec3} and Table~\ref{tab:sensitivity}; over the tested range, the mean deviation 
remains between $4.21\%$ and $6.09\%$, and the apparent improvement at larger $\lambda$ is 
shown to come at the cost of the $K^{*}$ sector rather than representing a uniformly better fit.

The spectrum derived from the effective radial Regge slope, $a=1.2519\ \mathrm{GeV}^2$ (Eq.~\ref{Eq:12}), and compared with soft-wall prediction of $4\kappa^2\approx1.09\ \mathrm{GeV}^2$. The elevation stems from the additional Coulombic and logarithmic terms included in the transverse potential. As noted in Sec.~\ref{sec4}, this value should be viewed as an effective slope obtained from a zero-intercept fit for comparison with Eq.~(\ref{Eq:10}), rather than as an exact parameter of the model. Unlike the analytically linear soft-wall potential, the modified potential yields transverse mass eigenvalues that deviate slightly from strict linearity in $(n+L+S/2)$, as shown in Tables 3, 4, and 5. Even so, the squared meson masses remain approximately linear in the radial quantum number,  consistent with an effective radial trajectory; with only three states per family ($n=0,1,2$),  this should be read as illustrative rather than a rigorous test of Regge behavior. The mass splitting in the strange sector continues to arise primarily from the longitudinal BdT contribution, since the transverse dynamics are flavor independent.

The largest remaining discrepancies are found for the excited pion states, where the deviations remain at the level of about $9$--$10\%$ (Table 3), somewhat larger than those of the vector mesons. This is consistent with the distinctive nature of the pion as a pseudo-Goldstone boson associated with spontaneous chiral symmetry breaking, an effect that is not explicitly incorporated into the present framework \cite{Choi2015,Li2022,Roberts2021,Ballon-Bayona2015,Brodsky2008}. This also indicates that radially excited pions include additional dynamics beyond those highlighted by a flavor-independent transverse potential \cite{Li2016}. Although the model does not include the full range of spin-dependent interactions or higher Fock-state contributions, it reproduces the main features of confinement and radial excitation with reasonable accuracy while maintaining a simple holographic description.

In summary, the present work demonstrates that a dimensionally consistent extension of the light-front holographic soft-wall potential, incorporating Coulombic and logarithmic interactions adapted from a heavy-meson light-front quark model through the instant-form/front-form correspondence of Ref.~\cite{Travinski2014}, leads to a noticeably improved description of the light-meson spectrum. The results also clarify the limitations of a flavor-independent treatment of the short-range interaction, suggesting that flavor-dependent extensions provide a natural direction for future investigation while preserving the underlying holographic framework.

\section{Conclusion}\label{sec7}

In this work, we investigated the radial excitation spectrum of light pseudoscalar and vector mesons within the framework of light-front holographic QCD. The transverse confining potential was extended by incorporating Coulombic and logarithmic interactions adapted from the heavy-meson light-front quark model of Ref.~\cite{Pandya2024}. Using the instant-form/front-form correspondence of Ref.~\cite{Travinski2014}, these additional terms were reformulated in a dimensionally consistent manner for the mass-squared eigenvalue equation, and the resulting light-front Schr\"odinger equation was solved numerically. With a single fitted parameter, $\alpha_C=0.172$, the extended potential reduces the mean deviation from the PDG masses for nine $\pi$, $\rho$, and $K^{*}$ states from $7.62\%$ in the baseline soft-wall model to $5.15\%$, corresponding to an overall improvement of approximately $32\%$, representing an in-sample reduction in mean deviation rather than an independent out-of-sample test (Sec.~\ref{sec5}).

The improvement is concentrated mainly in the pion and $\rho$ sectors, where the baseline soft-wall model systematically underestimates the excited-state masses. By comparison, the higher excited $K^{*}$ states are reproduced slightly less accurately than in the original model, reflecting the use of a single flavor-independent value of $\alpha_C$ for both the light and strange sectors. The largest remaining deviations are found for the excited pion states, consistent with the pion's unique role as a pseudo-Goldstone boson of spontaneously broken chiral symmetry, whose dynamics are not explicitly incorporated into the present framework. Despite these limitations, the modified potential holds the essential features of confinement, and the resulting squared meson masses continue to exhibit an approximately linear dependence on the radial quantum number. Although explicit spin-dependent interactions, higher Fock-state contributions, and flavor-dependent short-range interactions are not included, the model provides a simple, transparent, and quantitatively improved description of the light meson spectrum.

The results demonstrate that modest, dimensionally consistent extensions of the light-front holographic potential can significantly improve the phenomenological description of light mesons without modifying the underlying theoretical framework. Future refinements may include a flavor-dependent treatment of the short-range interaction to improve the strange sector, propagation of the uncertainty in $\kappa$ into the predicted mass spectrum, and extension of the formalism to other hadronic systems.

\section{Acknowledgement}

Abhisth Srivastava would like to acknowledge the support provided by Dr. Suneel Dutt, Dr. Harleen Dahiya and Dr. Arvind Kumar for their valuable insight during the study and preparation of manuscript. 

\appendix
\section{Numerical Method}
\label{secA1}

In this appendix, we describe the numerical procedure used to compute the transverse mass spectrum within light-front holographic QCD. The method solves the effective light-front Schr\"odinger equation for the transverse dynamics of a quark--antiquark system and determines the corresponding eigenvalues \cite{NumericalRecipes, GolubVanLoan} contributing to the meson mass.

\subsection{Transverse Light-Front Equation}

The starting point is the light-front Schr\"odinger equation for the transverse mode
\cite{Brodsky2015,Ahmady2021,Brodsky2009},
\begin{equation}
\left(-\frac{d^2}{d\zeta^2}
+\frac{4L^2-1}{4\zeta^2}
+U_\perp(\zeta)\right)\phi(\zeta)
=
M_\perp^2\,\phi(\zeta),
\label{Eq:A1}
\end{equation}
where $\zeta$ denotes the invariant transverse separation between the quark and antiquark, $L$ is the light-front orbital angular momentum, and $U_\perp(\zeta)$ is the effective confining potential, given here by the extended form of Eq.~(\ref{Eq:9}). The eigenvalue $M_\perp^2$ represents the transverse contribution to the squared meson mass. Throughout this work we consider only the $\pi$, $\rho$, and $K^{*}$ families, for which $L=0$, and the discussion below is restricted to this case.

\subsection{Exact Soft-Wall Basis}
\label{secA2}

Instead of discretizing Eq.~(\ref{Eq:A1}) on a spatial grid, we expand the solution in the exact eigenfunctions of the unperturbed ($\alpha_C=c_{\rm eff}=0$) soft-wall Hamiltonian,
\begin{equation}
\left(
-\frac{d^2}{d\zeta^2}
-\frac{1}{4\zeta^2}
+\kappa^4\zeta^2
\right)
\phi(\zeta)
=
\lambda\,\phi(\zeta).
\end{equation}

The normalized eigenfunctions are
\begin{equation}
\phi_n(\zeta)
=
\kappa\sqrt{2}\,
\zeta^{1/2}
e^{-\kappa^2\zeta^2/2}
L_n(\kappa^2\zeta^2),
\qquad
n=0,1,2,\ldots,
\label{Eq:A2}
\end{equation}
where $L_n$ denotes the Laguerre polynomial of degree $n$. These functions satisfy
\[
\int_0^\infty |\phi_n(\zeta)|^2\,d\zeta=1,
\qquad
\int_0^\infty
\phi_m(\zeta)\phi_n(\zeta)\,d\zeta
=\delta_{mn},
\]
with eigenvalues
\begin{equation}
\lambda_n
=
2\kappa^2(2n+1).
\label{Eq:A3}
\end{equation}

Together with the spin-dependent shift $2\kappa^2(J-1)$ in Eq.~(\ref{Eq:9}), these reproduce the exact soft-wall spectrum of Eq.~(\ref{Eq:10}),
\[
M^2=4\kappa^2(n+L+S/2),
\]
which serves as a benchmark for the numerical implementation. The eigenfunctions satisfy the required boundary conditions,
$\phi_n(\zeta)\rightarrow0$ as $\zeta\rightarrow0$ and
$\zeta\rightarrow\infty$, and form the basis used to solve the extended problem.

\subsection{Matrix Construction and Diagonalization}

The transverse wavefunction is expanded as
\begin{equation}
\phi(\zeta)
=
\sum_{n=0}^{N_{\rm basis}-1}
c_n\,\phi_n(\zeta),
\end{equation}
where the expansion is truncated at a finite basis size $N_{\rm basis}$.

Projecting Eq.~(\ref{Eq:A1}) onto this basis yields the matrix eigenvalue problem \cite{GolubVanLoan}
\begin{equation}
\sum_n
H_{mn}\,c_n
=
M_\perp^2\,c_m,
\qquad
H_{mn}
=
\lambda_n\delta_{mn}
+
\int_0^\infty
\phi_m(\zeta)
\left[
-\frac{4\alpha_C\kappa}{3\zeta}
+
c_{\rm eff}
\ln\!\left(\frac{\zeta}{\zeta_0}\right)
+
2\kappa^2(J-1)
\right]
\phi_n(\zeta)\,
d\zeta,
\label{Eq:A4}
\end{equation}
where the Coulombic and logarithmic interactions generate the off-diagonal matrix elements that couple different basis states. The integrals are evaluated numerically over
$\zeta\in(0,\zeta_{\rm max}]$, with
$\zeta_{\rm max}=40/\kappa$.
The resulting real symmetric matrix is diagonalized using standard linear algebra routines. Its eigenvalues correspond to the transverse mass spectrum
$M_\perp^2(n,L,S)$, while the eigenvectors determine the expansion coefficients of the corresponding transverse wavefunctions in the basis of Eq.~(\ref{Eq:A2}).

\subsection{Validation and Convergence}

The implementation was first validated by setting
$\alpha_C=c_{\rm eff}=0$, for which the numerical eigenvalues reproduce the analytic soft-wall spectrum of Eq.~(\ref{Eq:10}) to machine precision in both the pseudoscalar ($S=0$) and vector ($S=1$) sectors. This confirms that the basis expansion and numerical integration introduce no measurable error in the exactly solvable limit.

The convergence of the extended calculation was then examined by increasing the basis size $N_{\rm basis}$. Table~\ref{tab:convergence} lists the lowest three eigenvalues of the vector ($J=1$) sector for basis sizes ranging from 10 to 45. The calculated masses change by less than $0.4$ MeV between $N_{\rm basis}=25$ and $N_{\rm basis}=45$. The rapid convergence indicates that the Coulombic and logarithmic interactions act as relatively weak perturbations to the confining $\kappa^4\zeta^2$ potential, coupling each basis state primarily to nearby states. Consequently, a relatively small basis is sufficient to obtain converged eigenvalues. All results presented in this work were obtained with $N_{\rm basis}=25$.

\subsection{Summary}

The basis-expansion method described above provides an efficient and numerically stable approach for solving the extended light-front Schr\"odinger equation. Because the calculation is carried out in the exact soft-wall basis, the singular behavior at $\zeta=0$ is treated analytically rather than through spatial discretization, and the unperturbed spectrum is reproduced exactly before the additional interactions are introduced. Using the parameters listed in Table~\ref{tab:parameters}, the procedure is fully reproducible and can be extended straightforwardly to other bound-state systems within the same framework.

%Bibliography
\bibliographystyle{unsrt}  
\bibliography{sn-bibliography}

\end{document}